\documentclass[conference]{IEEEtran}
\IEEEoverridecommandlockouts
\usepackage{cite}
\usepackage{amsmath,amssymb,amsfonts}
\usepackage{graphicx}
\usepackage{booktabs}
\usepackage{multirow}
\usepackage{tikz}
\usetikzlibrary{arrows.meta,positioning,calc}
\usepackage[hidelinks]{hyperref}
\def\BibTeX{{\rm B\kern-.05em{\sc i\kern-.025em b}\kern-.08em T\kern-.1667em\lower.7ex\hbox{E}\kern-.125emX}}

\begin{document}

\title{A Cross-Domain Matrix Engine for Signal Processing and Deep Neural Net Inference }

\author{\IEEEauthorblockN{Mehdi Safarpour}
\IEEEauthorblockA{lastname.firstname[at]oulu.fi}}

\maketitle

\begin{abstract}
Matrix multiplication dominates both PHY signal processing and Deep Neural Network (DNN) inference, yet these domains place opposing demands on the matrix engine. Signal processing requires low-latency execution of relatively small matrix operations, whereas DNN inference, particularly at large batch sizes, prioritizes high throughput and energy efficiency for large matrix multiplications. While offering efficient memory access profile, a Systolic Array (SA) adds fill and drain delays per matrix. Those can be amortized for large matrices but not for small matrices. Conversely, a tensor unit completes the kernel in few cycles but would require repeated reads of same operand tiles over large matrices. This makes SA too slow for signal processing and tensor too inefficient for AI. We show that this need not be a choice between systolic and tensor structuring of Multiply–Accumulate (MAC) units. We propose a single engine that operates primarily as a SA and, through low-cost reconfiguration, transforms into a tensor-style engine on a per-operation basis. MUXes create both the spatial reduction adder tree and temporal reduction accumulators from the existing MACs and registers within the systolic array.
The proposed reconfigurable SA simulated using modified SCALE-Sim simulator, verified on an Zynq 702 FPGA and synthesized using ASAP 7nm FinFET library. SCALE-Sim simulations shows up to 7.5× lower latency for small complex value PHY kernels. Even on DNN workloads, it achieves up to 1.65× speedups over vanilla SA execution with trivial increase of memory accesses, retaining the systolic array's memory access advantage. Synthesis results estimate reconfigurable fabric costs only 8\% power and 13.6\% area overheads.

\end{abstract}


\section{Introduction}

Systolic Arrays (SA) are one of the most popular structures for implementation of matrix engines, especially for AI accelerators \cite{krishna2020data}. In mobile devices, both physical-layer baseband processing and neural-network inference are dominated by matrix arithmetic. Most DNN computations, including convolutional, fully connected, and attention layers, are commonly lowered to GEMM \cite{krishna2020data}. Similarly, wireless operations such as MIMO detection, rely heavily on matrix operations. Hence, supporting both domains on a shared engine becomes increasingly attractive\footnote{This work has been filed for a patent early 2026 \cite{safarpour2026reconfigurable}.}.

The two domains, however, place rather different demands on matrix engines. Wireless algorithms often involve serially processed relatively small matrices, under tight real-time constraints, e.g., iterative MIMO decoding composes of successive matrix operations whose results depend on previous computations. DNN workloads, on the other hand, generally involve larger matrices, where high throughput and efficient operand reuse are essential, as a large energy portion is spent on moving operands between the memory and compute engines \cite{krishna2020data}. A 2D systolic array \cite{krishna2020data} addresses the latter requirement well through its large matrix tile and substantial data reuse. However, its fill-and-drain delays and low utilization become significant for small or irregular matrices, where these overheads cannot be amortized. Tensor units make the opposite trade-off. By organizing MAC units to perform reductions in parallel \cite{tang2023performance}, they substantially reduce the latency of small matrix operations, but sacrifice operand reuse when processing large matrices. Repeated fetching of same tiles increases memory energy consumption \cite{krishna2020data}, as memory access can be up to O(10)$\times$ more expensive than arithmetic \cite{krishna2020data}.

\begin{figure}[t!]
\centering
\includegraphics[width=0.85\columnwidth]{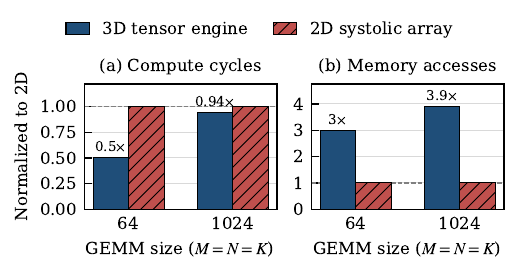}
\vspace{-15pt}
\caption{Compute cycles and memory accesses of tensor and systolic engines for small and large GEMMs. Tensor is with full memory bandwidth.}
\label{fig:motivaiton}
\vspace{-20pt}
\end{figure}

As illustrated in Fig.~\ref{fig:motivaiton}, the tensor engine offers substantial latency advantages for small GEMMs, while its advantage diminishes for larger matrices despite retaining considerably higher memory traffic. Neither organization is therefore efficient across both workload domains.

This paper proposes a reconfigurable matrix engine, depicted in Fig.~\ref{fig:regimes}, in which the same MAC units can operate either as a systolic array or as a tensor-style engine, without changing the peripheral memory system or interfaces. Unlike conventional systolic reshaping, the proposed organization changes how the reduction dimension is realized, allowing per-operation selection between low-latency tensor execution and high-reuse systolic execution. 

\section{Background and Related Work}
\vspace{-2pt}
In this section, a brief background on targeted applications, the utilized matrix engines, and related works is presented. 

\subsection{Application Domains}

\textit{Wireless physical layer:}
Many physical-layer algorithms reduce to small dense matrix operations. For example, linear massive MIMO detection \cite{hoydis2021} forms the Gram matrix ($\mathbf{G}=\mathbf{H}^{H}\mathbf{H}$), matched-filter vector ($\mathbf{H}^{H}\mathbf{y}$), followed by inversion or an approximation such as a Neumann series. All relatively small GEMM operations that are serialized. Precoding and channel estimation similarly rely on small sequential GEMMs. Complex-valued kernels can be mapped to real-valued GEMMs through the $[\mathrm{Re},-\mathrm{Im};\mathrm{Im},\mathrm{Re}]$ embedding. Compared with DNN workloads, wireless domain matrices are typically smaller, but must often require much tighter per operation latency constraints.

\begin{figure}[t!]
\centering
\includegraphics[width=\columnwidth]{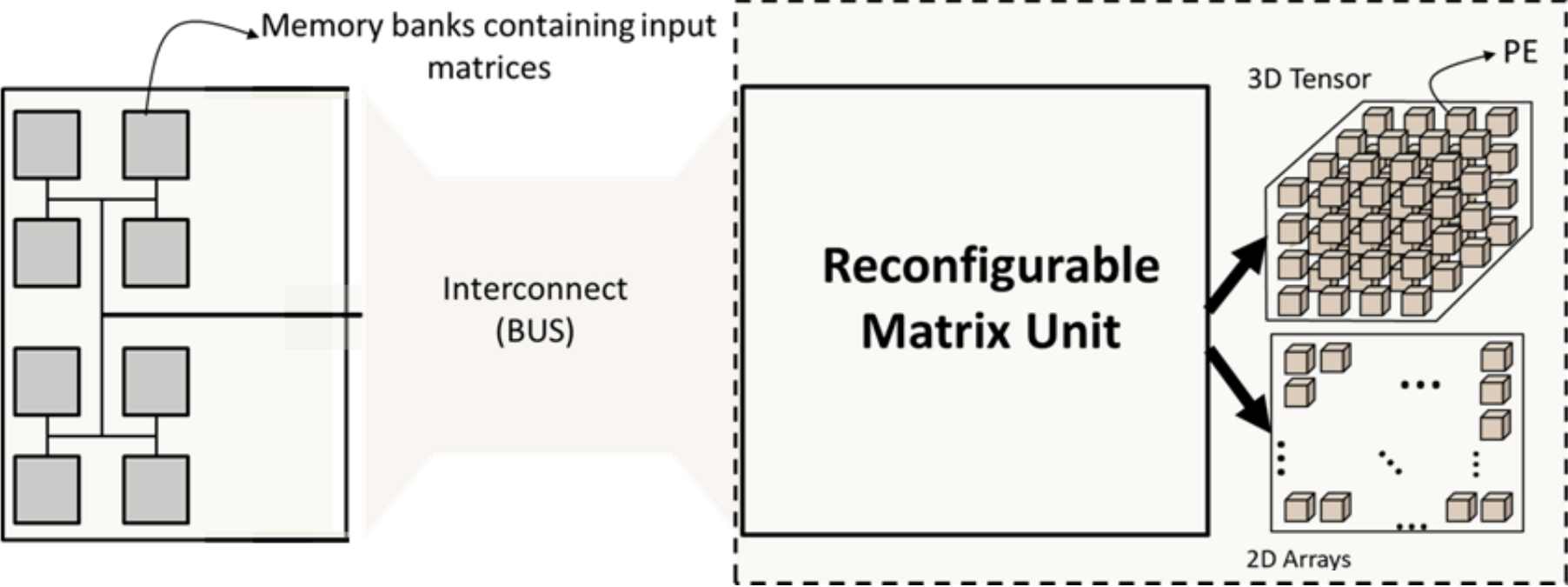}
\caption{The MAC units can be configured in tensor-style or and in 2D SA.}
\vspace{-10pt}
\label{fig:regimes}
\end{figure}

\textit{DNN inference:}
DNN inference is also dominated by matrix multiplication. Convolutional, fully connected layers and attention are commonly lowered to GEMM \cite{krishna2020data}. Matrix dimensions can vary substantially across layers and models. Batch size further changes the matrix shape, usually for benefit of higher throughput at the cost of higher latency. Thus, DNN workloads contain both large reuse-dominated GEMMs and smaller operations for which array utilization and latency become important.

\subsection{Matrix Engines}

\textit{Tensor units:}
Tensor engines such as NVIDIA Tensor Cores and the Ascend Cube \cite{tang2023performance} execute small dense matrix multiplications with parallelism across the reduction dimension. Instead of propagating operands through a 2-D wavefront as in SAs, multiple products along $K$ are generated concurrently and reduced through an adder network. This removes most of the systolic fill-and-drain latency, at the cost of smaller tile size (see Fig.~\ref{fig:motivaiton}). Total multiplications and adders count is equal to that of an SA. Adder performs spatial reduction, and temporal reduction based an an output-stationary (OS) dataflow.

\textit{Systolic Arrays:}
In this work we consider OS systolic array, while the results are applicable to weight stationary and input stationary flavors as well. In an $R\times C$ output-stationary systolic array, each processing element accumulates one output while operands propagate along the rows and columns. Each GEMM tile incurs a fill and drain overhead in addition to computation. This is amortized for large GEMMs, i.e., with large reduction, but becomes significant when the mapped dimensions are small relative to the array. 

\subsection{Related works: Reconfigurable Systolic Arrays}

Several architectures have addressed the utilization loss of fixed-size systolic arrays by reconfiguring their shape or dataflow. FlexSA \cite{lym2020flexsa} partitions a large systolic array into smaller parallel sub-arrays, while ReDas \cite{han2024redas} enables finer-grained reshaping and multiple systolic dataflows through additional PE routing. However, these approaches are less effective for small, sequentially dependent GEMMs, common in wireless algorithms, where limited parallelism leaves many MAC units underutilized. Our approach reconfigures the reduction dimension itself, enabling low-latency execution of individual small GEMMs, not just improve utilization.

\begin{figure}[t]
\centering
\includegraphics[width=\columnwidth]{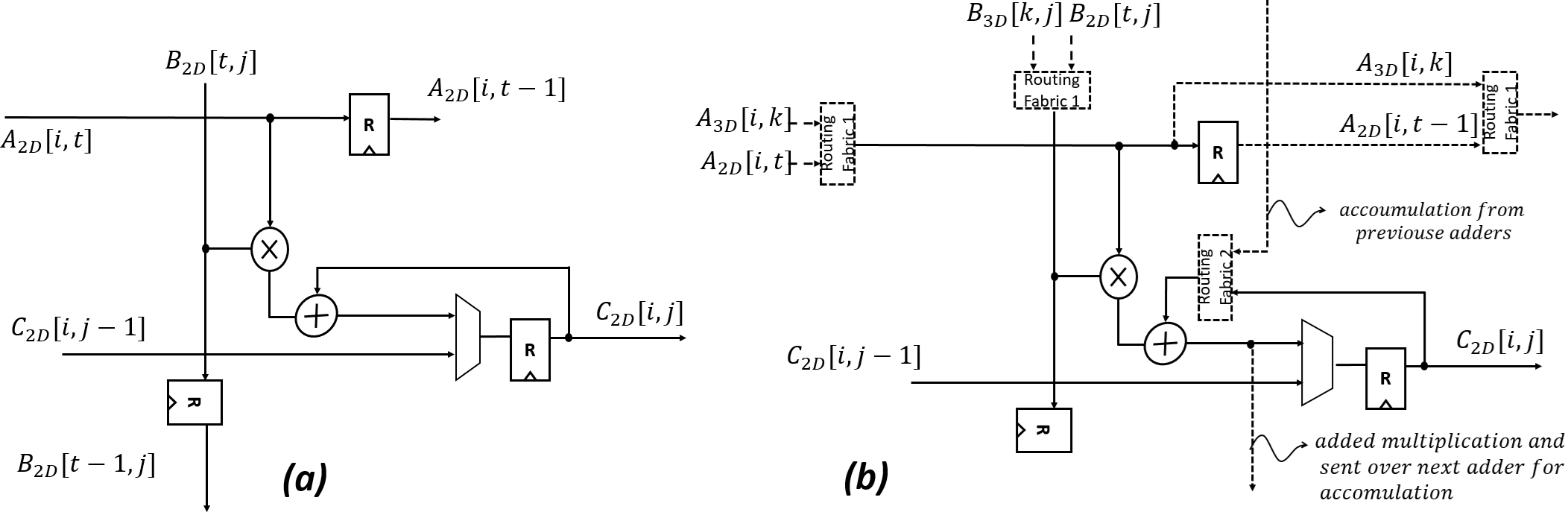}
\caption{The internal routing in PE of SA (a) requires just few reconfiguration fabric (as MUXes or transmission gates) for adder tree of a Tensor core (b).}
\vspace{-10pt}
\label{fig:PEs}
\end{figure}

\section{Reconfigurable Matrix Engine}

\label{sec:arch}

The same MAC units of a systolic array (SA) can either form a large 2D systolic plane or be reorganized into smaller planes with parallelism along the reduction dimension $K$, forming a tensor-style so called 3D engine \cite{tang2023performance}. Unlike conventional systolic reshaping, the proposed reconfiguration changes how the reduction dimension is realized by forming a spatial reduction tree from existing SA resources. 

In our configuration, a $64\times64$ SA is reconfigured into a $16\times16\times16$ tensor engine using the same 4096 MACs. The SA accepts 64 elements per operand per cycle, whereas a tensor operation requires two $16\times16$ operand tiles, supplied over four cycles through the shared interface. The SA's skew buffer is reused for tensor operand buffering, while existing accumulators on SA form the spatial and temporal reduction tree. For complex-valued PHY operations, local complex-to-real expansion further reduces operand loading to two cycles.

Restricting tensor mode to the SA bandwidth requires minimal hardware modifications but limits MAC utilization compared to an unrestricted tensor engine. Nonetheless, even with this restriction, tensor mode substantially reduces the latency of small kernels, particularly complex-valued operations.

\subsubsection{Organization}

\begin{figure}[t]
\centering
\includegraphics[width=\columnwidth]{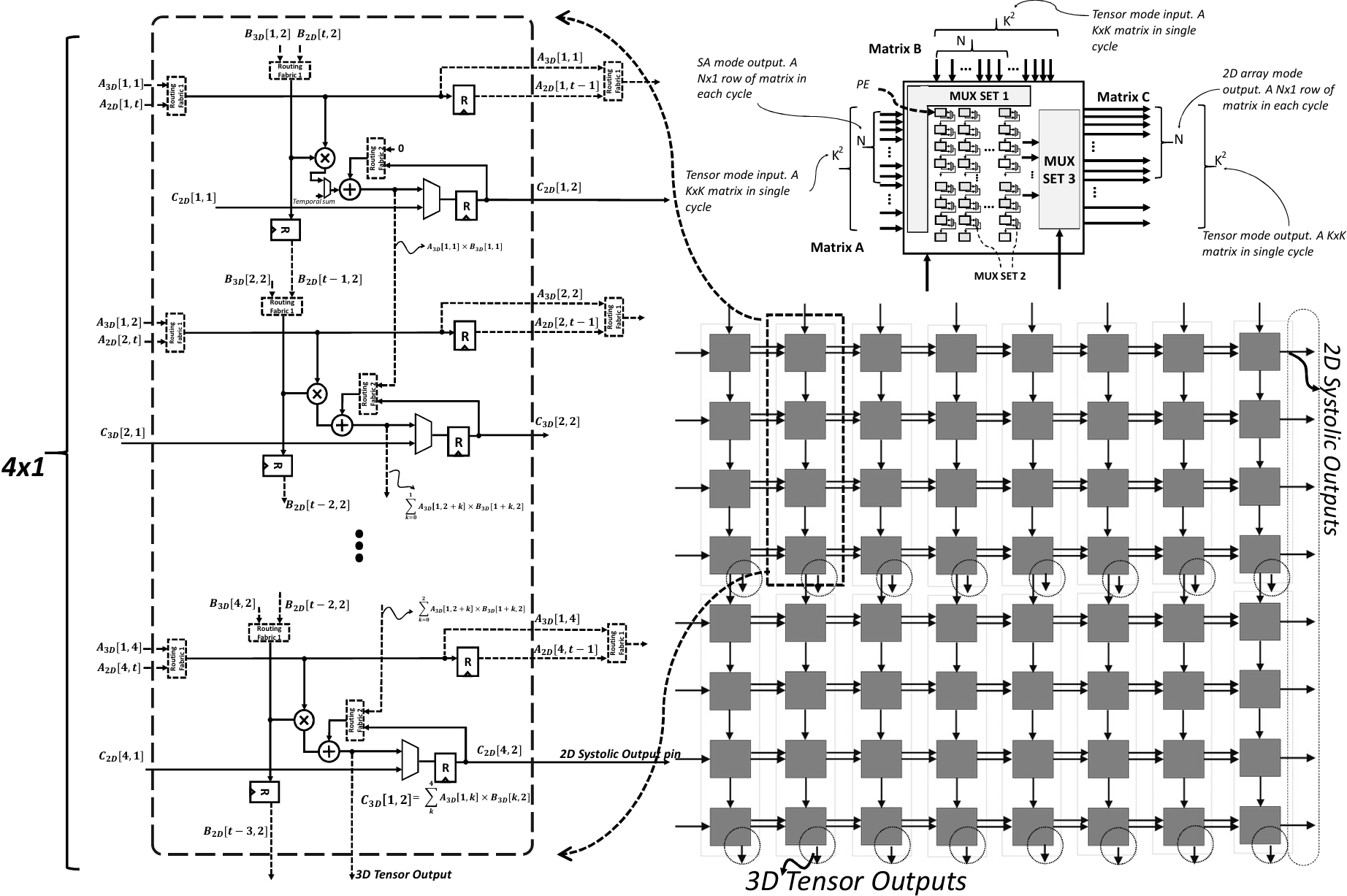}
\vspace{-10pt}
\caption{Top view of integration of systolic and tensor cores as reconfigurable engine (64 MAC example) with MUXes shows as reconfiguration fabric.}
\vspace{-15pt}
\label{fig:regimes2Dto3D}
\end{figure}

Figures~\ref{fig:regimes},\ref{fig:PEs} and \ref{fig:regimes2Dto3D} depict the two organizations using an $8\times8$ SA reconfigurable into a $4\times4\times4$ tensor engine for illustration simplicity. In the actual implementation, the $64\times64$ SA is reorganized into sixteen $16\times16$ planes along the reduction dimension $K$. Each plane receives the corresponding column of $\mathbf{A}$ and row of $\mathbf{B}$, generating products $a_{ik}b_{kj}$ that are summed through a spatial reduction tree.

As illustrated in Figs.~\ref{fig:PEs} and \ref{fig:regimes2Dto3D}, reconfiguration is achieved through three sets of routing fabrics, shown in Fig.~\ref{fig:regimes2Dto3D} top right. Notice the depiction assumes a $N\times N$ SA is reconfigured to $K \times K \times K$ tensor. Notice the $N$ and $K$ here refer to physical array's dimensions not matrix sizes. Our implementation restricts both modes to SA's interface.  Routing fabrics are mainly 2:1 multiplexers here. The first set controls operand routing, switching between systolic streaming and tensor-style broadcasting. The second changes the accumulation structure, i.e., in systolic mode, each PE accumulates partial results over time using its adder-register pair, whereas in tensor mode, the existing adders are interconnected to form a spatial reduction tree across the planes, see Fig~\ref{fig:PEs}. The third controls output routing, as systolic results leave the array row by row, while tensor mode produces a complete $16\times16$ output tile that is transferred through the shared output interface over multiple cycles. All routing configurations are controlled by a single mode bit distributed across the array.

\subsubsection{Timing, memory access and shared interface}
Both modes share the same memory interface, with bandwidth determined by the systolic array. Consequently, tensor mode must load its operand tiles over multiple cycles rather than in a single cycle, limiting its performance compared to an unrestricted tensor engine. Nevertheless, by exploiting parallelism along the reduction dimension, tensor mode substantially reduces the execution latency of small GEMMs despite the shared bandwidth constraints. For an $M\times N\times K$ GEMM, cycles each engine take are:
\begin{align}
T_{\mathrm{sys}} &= \Big\lceil \tfrac{M}{64}\Big\rceil \Big\lceil \tfrac{N}{64}\Big\rceil (K + 126), \label{eq:tsys}\\
T_{\mathrm{ten-limited}} &= \Big\lceil \tfrac{M}{16}\Big\rceil \Big\lceil \tfrac{N}{16}\Big\rceil \Big(4\Big\lceil \tfrac{K}{16}\Big\rceil + 3 + 4\Big).
\end{align}


\subsubsection{Operand traffic}
Each engine reads valid operand elements only so edge padding is not fetched and invalid outputs are not written. The SA reads every $\mathbf{A}$ element once per column fold and every $\mathbf{B}$ element once per row fold and tensor mode with 16-element granularity:
\begin{align}
A_{\mathrm{sys}} &= MK\Big\lceil \tfrac{N}{64}\Big\rceil + KN\Big\lceil \tfrac{M}{64}\Big\rceil + MN, \label{eq:asys}\\
A_{\mathrm{ten}} &= MK\Big\lceil \tfrac{N}{16}\Big\rceil + KN\Big\lceil \tfrac{M}{16}\Big\rceil + MN. \label{eq:aten}
\end{align}
Since $\lceil x/16\rceil\ge\lceil x/64\rceil$, tensor mode never moves fewer operands than systolic mode. Tensor approaches $4\times$ when both dimensions are large. Equations (1) to (5) show why one structure cannot serve both application domains.

\subsubsection{Mode selection}
The compiler or scheduler selects the mode per operations known at compile time or at dispatch. The selection policy in this paper is to take the mode with fewer cycles and if the engine is within 10\%, take whichever has fewer accesses by Eq.~\eqref{eq:asys} and \eqref{eq:aten} and the assume one cycle per transition after pipeline completion.

Finally, for complex matrices, the restricted bandwidth of the SA in tensor mode can be mitigated, by generating the larger real matrices locally from the real and imaginary components of the complex inputs. This small modification is particularly useful for wireless PHY processing, which predominantly involves complex-valued matrices. This reduces number of cycles to load a new matrix to halve: \begin{align}T_{\mathrm{ten-limited}} &= \Big\lceil \tfrac{M}{16}\Big\rceil \Big\lceil \tfrac{N}{16}\Big\rceil \Big(\textbf{2}\Big\lceil \tfrac{K}{16}\Big\rceil + 3 + \textbf{2}\Big).
\label{complex_value_loads}
\end{align}

 \begin{figure}[t]
\centering
\includegraphics[width=0.85\columnwidth]{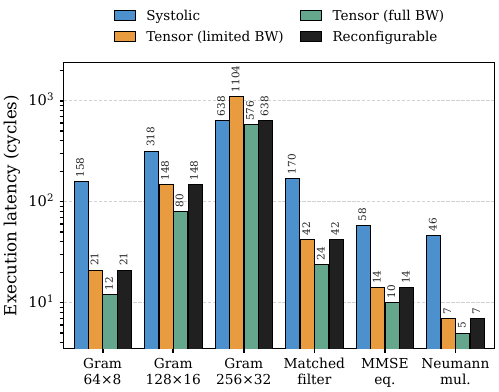}
\vspace{-5pt}
\caption{Complex wireless PHY matrix
operations. The lowest latency mode is selected for reconfigurable. SA does not perfom well for PHY.}
\vspace{-15pt}
\label{fig:telecom}
\end{figure}

 \section{Methodology}
\label{sec:method}

To measure performance across different workloads, the SCALE-Sim simulator \cite{raj2025scale} was used. Workloads consists of common telecom jobs and various DNN models. Also, to estimate the routing overhead introduced by reconfiguration, a conventional tensor core, a systolic array, and the proposed reconfigurable array were described in HDL and synthesized and functionality verified.

All cycle and access counts are produced by SCALE-Sim. Systolic mode uses the stock output-stationary SA model with a 64$\times$64 array, 64KB operand and output scratchpads, and the default bandwidth estimation memory model \cite{raj2025scale}. Tensor mode is added as a new compute engine such that it consumes matrices for a 16$\times$16$\times$16 tensor with a four cycle pipeline, while the unmodified scratchpad, DRAM prefetch and stall machinery serves all investigated organizations \cite{tang2023performance}, hence, memory interconnect bandwidth is assumed to be limited to that of SA. Valid SRAM operand reads and output writes are counted and register accesses are ignored. In systolic mode, edge tiles do not retain their full cycle cost and idle lanes do not issue null requests as it is common in practice. This emulates behavior of mentioned reconfigurable SAs, e.g., \cite{lym2020flexsa}. Analytical cycles may match with SCALE-Sim's report when there are no memory stalls due to small size of problem.

The selected wireless and AI workloads represent different requirements on the matrix engine. Wireless kernels assume 64 antennas and 8 users unless otherwise stated. Various DNN workloads, lowered according to SCALE-Sim's convention ($M$ output pixels, $N$ filters, and $K$ receptive field), were selected to demonstrate that the solution also benefits conventional AI workloads while retaining its efficiency for large matrix operations.

\section{Evaluation and Discussion}
\label{sec:eval}
Performance and memory accesses were evaluated using SCALE-Sim. The design was first functionally verified on an FPGA with results of synthesis in a 7-nm process reported.

\begin{figure}[t]
\centering
\includegraphics[width=0.99\columnwidth]{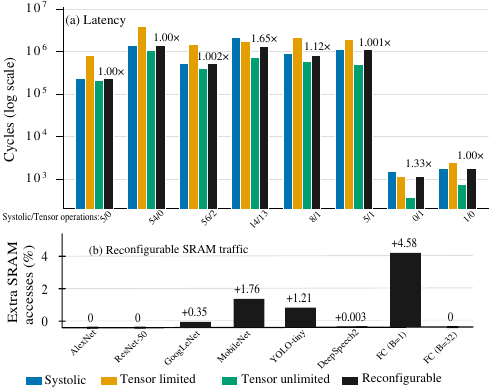}
\vspace{-15pt}
\caption{AI GEMM workloads on different engines. Share of SA and tensor operation are noted as Systolic/Tenser.}
\vspace{-15pt}
\label{fig:AI_workload}
\end{figure}

\subsubsection{Wireless kernels}

\label{sec:telecom}
Figure~\ref{fig:telecom} presents per-kernel results under the same memory-interface constraint. For latency-critical wireless PHY operations, tensor mode reduces latency by $2.15–7.52\times$. However, for larger matrices, systolic execution remains faster which of course is due to memory-interface restriction on tensor. The reconfigurable is the lowest of two modes. However, limited bandwidth tensor remain competitive to unlimited bandwidth tensor version for complex-valued operations, as local complex-to-real expansion reduces input and output transfers from four to two cycles each, further improving tensor-mode efficiency under limited bandwidth. The reported results cover only matrix multiplications.

\subsubsection{Neural networks}
\label{sec:ai}
Figure~\ref{fig:AI_workload} presents topology-level results. Although the primary motivation of this work is to enable SAs to meet the latency demands of wireless kernels, the results interestingly show that per-operation reconfiguration also improves DNN performance over conventional SAs, with only a small increase in SRAM accesses for most workloads. The benefits are particularly evident in MobileNet, YOLO-tiny, and low batch Fully Connected (FC). For large batch size SA turns out to benefit over tensor and hence reconfigurable array is kept configured to SA mode. For larger, convolution-dominated workloads, such as ResNet-50, the policy retains SA execution, preserving performance and memory efficiency. 

\subsubsection{Synthesis Results}

The design first was verified functionally on a Zynq SoC FPGA then synthesized for 8-bit integer \cite{clark2016asap7} using OpenROAD tool (see Fig.~7). Note, the numbers only include datapath and not memory access.

\subsubsection{Discussion}
The results in Fig.~\ref{fig:telecom} and Fig.~\ref{fig:AI_workload} reflect an example mode selection policy that can be tuned to favor performance or energy efficiency. For overheads, note all engines use INT8 arithmetic; higher-precision and multiple data type support substantially reduce the relative overheads. Assuming 5~pJ/byte access cost to a 64KB SRAM \cite{krishna2020data}, SA consumes $\approx 2\times$ less total energy than the tensor per ResNet-50 inference. Reconfigurable maintains this advantage while enabling low latency of the tensor.
\begin{figure}[t]
    \centering

    \begin{minipage}[t]{0.32\columnwidth}
        \vspace{0pt}
        \centering
        \includegraphics[width=92pt]{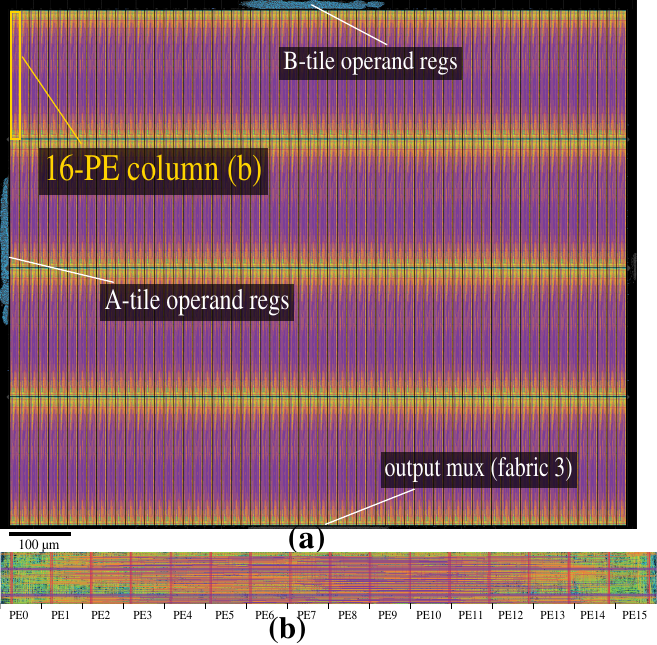}

        \vspace{1mm}
        \small
        \scriptsize{(a) 64 × 4 grid of 16-PE columns. (b) one 16-PE column (rotated). }
    \end{minipage}
    \hfill
    \begin{minipage}[t]{0.65\columnwidth}

        \vspace{0pt}
        \scriptsize
        \begin{tabular}{lccc}
            \hline
            \scriptsize
            &  SA &   Tensor$^{\mathrm{\triangle}}$ & Reconf.  \\
            \hline
            Technology
                & \multicolumn{3}{c}{ASAP7 7-nm FinFET} \\

            Supply voltage
                & \multicolumn{3}{c}{0.70~V} \\

            MACs
                & \multicolumn{3}{c}{4096} \\

            Freq.
                & \multicolumn{3}{c}{500~MHz} \\

            Area (mm$^2$)
               & 0.308 & 0.167 & 0.350   \\
               
            Power (mW)
               & 848 & 417 & 915   \\
               
            Overhead\tiny{(Area)} $^{\mathrm{\otimes}}$ 
                & -- & -- & $13.6\%$   \\
            
            Overhead\tiny{(Power)} $^{\mathrm{\otimes}}$ 
                & -- & -- & $8\%$   \\




            Tile size
               & 64$\times$64  & 16$\times$16 &    Variable \\

            Mem. BW$^*$
              & 192 & 768 & 192   \\
            \hline
        \end{tabular}

        \vspace{0.5mm}
        \small
        \vspace{0.5mm}
        \scriptsize{(c) Synthesis reports.}\\ $^*$ \tiny{\textit{Elements per cycle.}}~~~~~~$^\otimes$\tiny{Reconfigurable SA-mode versus base SA.} \\$^\triangle$ \tiny{\textit{With unlimited full memory BW.}}
    \end{minipage}

\caption{Synthesis results of the reconfigurable engine.}
    \vspace{-10pt}
\end{figure}

\section{Conclusion}

The proposed reconfigurable engine combines the advantages of both organizations, retaining the SA's memory efficiency for large AI workloads while providing low-latency tensor execution for small PHY operations. With modest hardware overhead, it enables systolic-array-based accelerators in mobile devices to support both AI inference and wireless baseband processing on a shared compute engine. It also benefits low-batch-size AI inference, extending the applicability to a wider range of edge AI applications.

\bibliographystyle{IEEEtran}
\bibliography{refs}

@article{clark2016asap7,
  title={ASAP7: A 7-nm finFET predictive process design kit},
  author={Clark, Lawrence T and Vashishtha, Vinay and Shifren, Lucian and Gujja, Aditya and Sinha, Saurabh and Cline, Brian and Ramamurthy, Chandarasekaran and Yeric, Greg},
  journal={Microelectronics Journal},
  volume={53},
  pages={105--115},
  year={2016},
  publisher={Elsevier}
}

@article{lym2020flexsa,
  title={FlexSA: Flexible systolic array architecture for efficient pruned DNN model training},
  author={Lym, Sangkug and Erez, Mattan},
  journal={arXiv preprint arXiv:2004.13027},
  year={2020}
}

@article{han2024redas,
  title={Redas: A lightweight architecture for supporting fine-grained reshaping and multiple dataflows on systolic array},
  author={Han, Meng and Wang, Liang and Xiao, Limin and Cai, Tianhao and Wang, Zeyu and Xu, Xiangrong and Zhang, Chenhao},
  journal={IEEE Transactions on Computers},
  volume={73},
  number={8},
  pages={1997--2011},
  year={2024},
  publisher={IEEE}
}

@patent{safarpour2026reconfigurable,
  author       = {Safarpour, Mehdi and Dicander, Hakan},
  title        = {Reconfigurable Processing Elements for Matrix Multiplication},
  year         = {2026},
    note         = {Filed Feb. 2026. (Pending patent)},
}

@article{hoydis2021,
  author={Hoydis, Jakob and Aoudia, Fay\c{c}al Ait and Alkhateeb, Ahmad and Cammerer, Sebastian and others},
  title={Toward a {6G} {AI}-native air interface},
  journal={IEEE Commun. Mag.},
  volume={59}, number={5}, pages={76--81}, year={2021}}

@article{tang2023performance,
  title={Performance modeling on DaVinci AI core},
  author={Tang, Yifeng and Wang, Cho-li},
  journal={Journal of Parallel and Distributed Computing},
  volume={175},
  pages={134--149},
  year={2023},
  publisher={Elsevier}
}

@article{krishna2020data,
  title={Data orchestration in deep learning accelerators},
  author={Krishna, Tushar and Kwon, Hyoukjun and Parashar, Angshuman and Pellauer, Michael and Samajdar, Ananda},
  year={2020},
  publisher={Springer}
}

@inproceedings{raj2025scale,
  title={SCALE-Sim v3: A modular cycle-accurate systolic accelerator simulator for end-to-end system analysis},
  author={Raj, Ritik and Banerjee, Sarbartha and Chandra, Nikhil and Wan, Zishen and Tong, Jianming and Samajdhar, Ananda and Krishna, Tushar},
  booktitle={2025 IEEE International Symposium on Performance Analysis of Systems and Software (ISPASS)},
  pages={186--200},
  year={2025},
  organization={IEEE}
}

\end{document}